\documentclass[usenatbib]{mn2e}

\pdfoutput=1

\usepackage{fixltx2e}
\usepackage{mathptmx}
\usepackage[pdftex]{graphicx}

\newcommand{\Mpc}{\rm\thinspace Mpc}

\newcommand{\km}{\rm\thinspace km}

\newcommand{\cm}{\rm\thinspace cm}

\newcommand{\pcmcu}{\hbox{$\cm^{-3}\,$}}

\newcommand{\Gyr}{\rm\thinspace Gyr}
\newcommand{\s}{\rm\thinspace s}

\newcommand{\keV}{\rm\thinspace keV}

\newcommand{\erg}{\rm\thinspace erg}

\newcommand{\ergps}{\hbox{$\erg\s^{-1}\,$}}

\newcommand{\kmps}{\hbox{$\km\s^{-1}\,$}}

\newcommand{\kmpspMpc}{\hbox{$\kmps\Mpc^{-1}$}}

\newcommand{\Zsun}{\hbox{$\thinspace \mathrm{Z}_{\odot}$}}

\begin{document}

\title{Sound waves in the intracluster medium of the Centaurus cluster}
\author[Sanders \& Fabian]{J.S. Sanders\thanks{E-mail: jss@ast.cam.ac.uk} and
    A.C. Fabian\\
    Institute of Astronomy, Madingley Road, Cambridge. CB3 0HA}
\maketitle
  
\begin{abstract}
  We report the discovery of ripple-like X-ray surface brightness
  oscillations in the core of the Centaurus cluster of galaxies, found
  with 200~ks of \emph{Chandra} observations.  The features are
  between 3 to 5 per cent variations in surface brightness with a
  wavelength of around 9~kpc. If, as has been conjectured for the
  Perseus cluster, these are sound waves generated by the repetitive
  inflation of central radio bubbles, they represent around $5\times
  10^{42}\ergps$ of spherical sound-wave power at a radius of
  30~kpc. The period of the waves would be $10^7$~yr. If their power
  is dissipated in the core of the cluster, it would balance much of
  the radiative cooling by X-ray emission, which is around $1.3 \times
  10^{43} \ergps$ within the inner 30~kpc. The power of the sound
  waves would be a factor of four smaller that the heating power of
  the central radio bubbles, which means that energy is converted into
  sound waves efficiently.
\end{abstract}


\section{Introduction}
Ripples in the X-ray surface brightness of galaxy clusters were first
discovered in the Perseus cluster of galaxies \citep{FabianPer03}. As
the surface brightness is roughly proportional to the density of the
intracluster medium (ICM) squared, these are density
ripples.

\cite{FabianPer03} explained these variations as sound waves generated
by the inflation of the bubbles of relativistic plasma in the core of
the cluster by the central active nucleus. Providing that these sound
waves can be dissipated as they travel \citep{FabianReynolds05}, they
have the ability to transport energy from the central nucleus to the
core of the cluster. If they carry significant quantities of energy,
they may provide the distributed heating required \citep{Voigt04} to
prevent large quantities of the ICM from cooling
\citep{PetersonFabian06,McNamaraNulsen07}.  Subsequent theoretical
work \citep{Ruszkowski04,Sijacki06} have shown that sound waves may be
able to help heat the cores of clusters in a gentle distributed fashion.

A deep 900~ks observation of Perseus showed the ripples in exquisite
detail enabling their amplitude to be measured
\citep{FabianPer06}. Subsequently, we used the amplitude of the
ripples to calculate the amount of energy propagated in them if they
are sound waves \citep{SandersPer07}. They would carry enough energy
in Perseus to combat a significant fraction of the cooling rate, and
appear to decline in power with radius.

The Perseus observations also reveal thick spherical shells around
each inner bubble at a higher gas density and pressure than their
surroundings. These high pressure shells have a sharp outer edge where
the density abruptly jumps by about 30 per cent, and is interpreted as
a weak shock. We have assumed that such high pressure regions
propagate outward to become the ripples. A problem with the weak shock
interpretation is that there is no temperature jump coincident with
the density one. The observed limit on any jump in temperature is
inconsistent with an adiabatic shock.  The region has been studied and
discussed in detail by \cite{Graham08Per}. It is noted there that the
Northern optical filaments of cold gas stop at the shock front, so
cold gas is likely being mixed with the hot gas there due to
turbulence generated at that front. This could explain a reduction in
temperature of the post shock gas.

Sound waves or ripples have not been detected in other
systems. However, they may still be an important contribution to the
heating processes in clusters as they could be difficult to
detect. They were not found until 200~ks of \emph{Chandra} time of the
Perseus cluster was obtained. Perseus is the X-ray brightest galaxy
cluster by a factor of 1.7 (in the 2-10~keV band;
\citealt{Edge90}). Clearly the detectability of ripples will vary with
cluster surface brightness and observation time. It will also vary
with ripple amplitude and wavelength and with cluster
properties. These effects have been examined numerically by
\cite{Graham08}, finding that it is difficult to observe these
features with current instruments in clusters other than Perseus. The
detectability of ripples, however, depends on unknown factors, such as
the wavelength and amplitude of the ripples. These factors can be
estimated with a large degree of uncertainty from the properties of
the cluster, assuming that they are responsible for heating the
cluster core and have a wavelength similar to the bubble size.

Here we report on some ripples which we find in a 200~ks
\emph{Chandra} observation of the Centaurus cluster of galaxies. These
data were examined previously by \cite{Fabian05} and
\cite{SandersEnrich06}, examining the thermal distribution of gas in
the cluster core and its metallicity.

The Centaurus cluster of galaxies is a nearby ($z=0.0104$;
\citealt{LuceyCurrieDickens86a}) X-ray bright galaxy cluster
($L_{X,2-10\keV} = 2.9 \times 10^{43} \ergps$; \citealt{Edge90}). It
contains a low radiative efficiency nucleus \citep{Taylor06} and
distinct bubbles of relativistic plasma \citep{Taylor02} displacing
the intracluster medium, as seen by \emph{Chandra}
\citep{SandersCent02}.

We assume $H_0 = 70 \kmpspMpc$, which translates into scale of 213~pc
per arcsec.

\section{Data analysis}
We obtained \emph{Chandra} observations with IDs 504, 4954, 4955 and
5310. Each of these observations used the ACIS-S detector aimed at the
central X-ray peak of Centaurus.  Exposure maps for each of the
observations and for each of the detector CCDs were created with the
\textsc{ciao} \textsc{mkexpmap} tool, binning the detector pixels by a
factor of 2 and leaving the observation time factor in the map by
disabling normalization. We used the appropriate bad pixel maps and
attitude files to create these maps and assumed a monochromatic energy
of 1.5~keV. We created images of Centaurus for each observation and
detector between 0.5 and 6~keV from the event files with the same
spatial binning. We added all the images together to make a total
image, and all the exposure maps to make a total exposure map. A total
exposure-corrected image was created by dividing the total image by
total exposure map.

\subsection{Fourier filtering}
To search for features in the exposure-corrected image we used a Fast
Fourier Transform (FFT) high-pass filter as in
\cite{SandersPer07}. The filter removes smooth underlying cluster
emission and so is similar to the unsharp-mask method.  Firstly, we
truncated the exposure-corrected values in the image to reduce the
level of the bright emission in the cluster centre. This was done
because using Fourier filters on bright peaks can introduce `fringes'
on the filtered image, which may look like additional ripples. We
tested that the features in the image were robust with different
clipping and filtering parameters. We also removed point sources from
the image before filtering, by selecting them by eye, then replacing
the source pixels with values selected randomly from the values of the
surrounding pixels.

\begin{figure*}
  \includegraphics[width=0.7\textwidth]{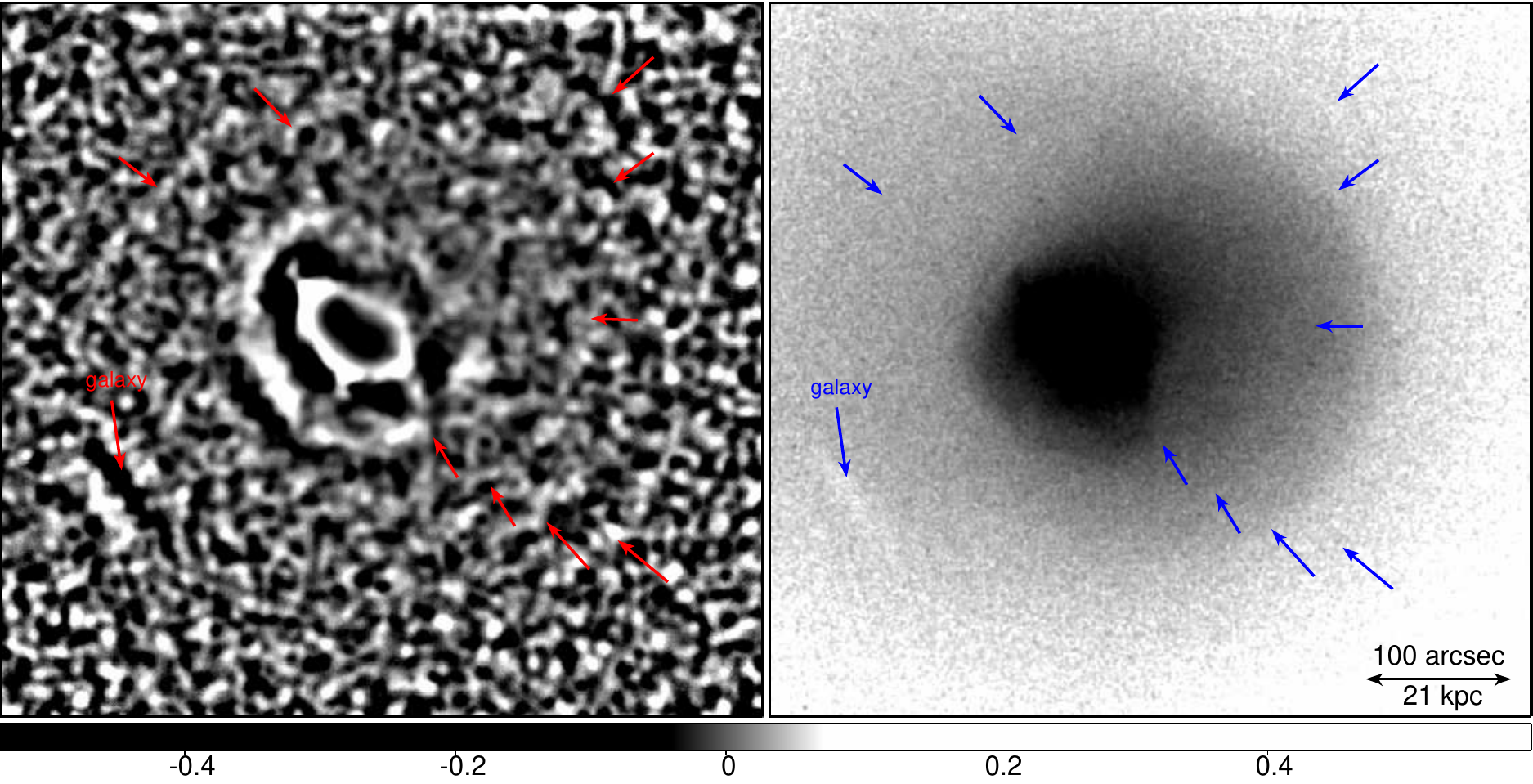}
  \caption{(Left) Ratio of FFT low-pass filtered image to
    Gaussian-smoothed image (20 arcsec). (Right) Exposure-corrected
    image, smoothed with a Gaussian with $\sigma = 2$~arcsec. Point
    sources were removed and replaced with random values from their
    surrounding pixels. We mark strong features with arrows at the
    same positions on the two panels.}
  \label{fig:fftimage}
\end{figure*}

In our analysis we removed spatial frequencies longer than 10 per cent
of the size of the image ($800\times0.98$ arcsec pixels square),
allowing through all frequencies shorter than 5 per cent of the size
of the image. We allowed intermediate frequencies through the filter
with a linear filter between these two threshold spatial
frequencies. We then computed the ratio between the filtered map and
the original image, after smoothing by a 20 arcsec Gaussian, to allow
the features in the centre and outskirts to be seen. The filtered
image and a smoothed unfiltered image are shown in
Fig.~\ref{fig:fftimage}.

\subsection{Surface brightness profiles}
\label{sect:profiles}
\begin{figure}
  \centering
  \includegraphics[width=0.72\columnwidth]{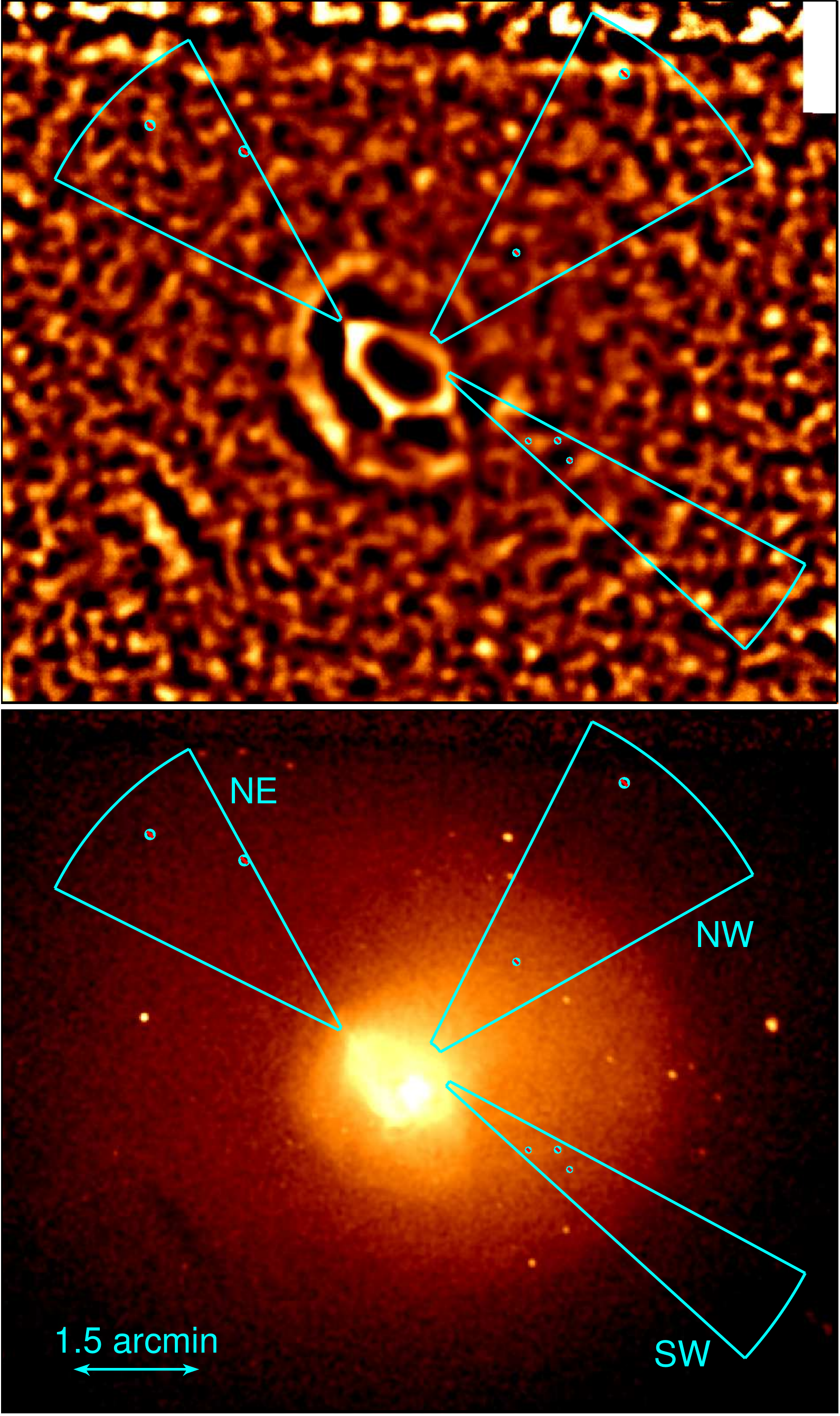}
  \caption{Regions examined in surface brightness profiles. The top
    panel is the ratio of the FFT-filtered image, smoothed by a
    Gaussian of 9~arcsec. The bottom is the exposure-corrected image
    smoothed by 3~arcsec. Also shown are the excluded point
    sources. North is upwards in these images and east is to the
    left.}
  \label{fig:profregions}
\end{figure}

To verify that the features that we have found are not artifacts
produced by the Fourier filtering technique and to measure the
amplitude of the features, we created surface brightness profiles in
different sectors. In Fig.~\ref{fig:profregions} we show the sectors
used to make the profiles. Note that the sectors are not centred on
the active nucleus, but are centred so that the surface brightness
features are aligned with the radial bins.

To produce a profile we take a sector and split it into radial
bins. From each bin we count the number of counts from each of the
separate observations in the 0.5 to 6~keV band. We also take the
average value of the total exposure map in each bin (single-pixel
binned exposure maps are used in this part of the analysis) and
multiply by the area of the bin in pixels. We divide the total number
of counts by the total exposure map value to make a surface brightness
measurement. Point sources were removed from the regions by eye. Note
that as the profiles are centred to match the ripples, the radii in
the profiles do not correspond to the same radius in the cluster.

\begin{figure}
  \includegraphics[width=\columnwidth]{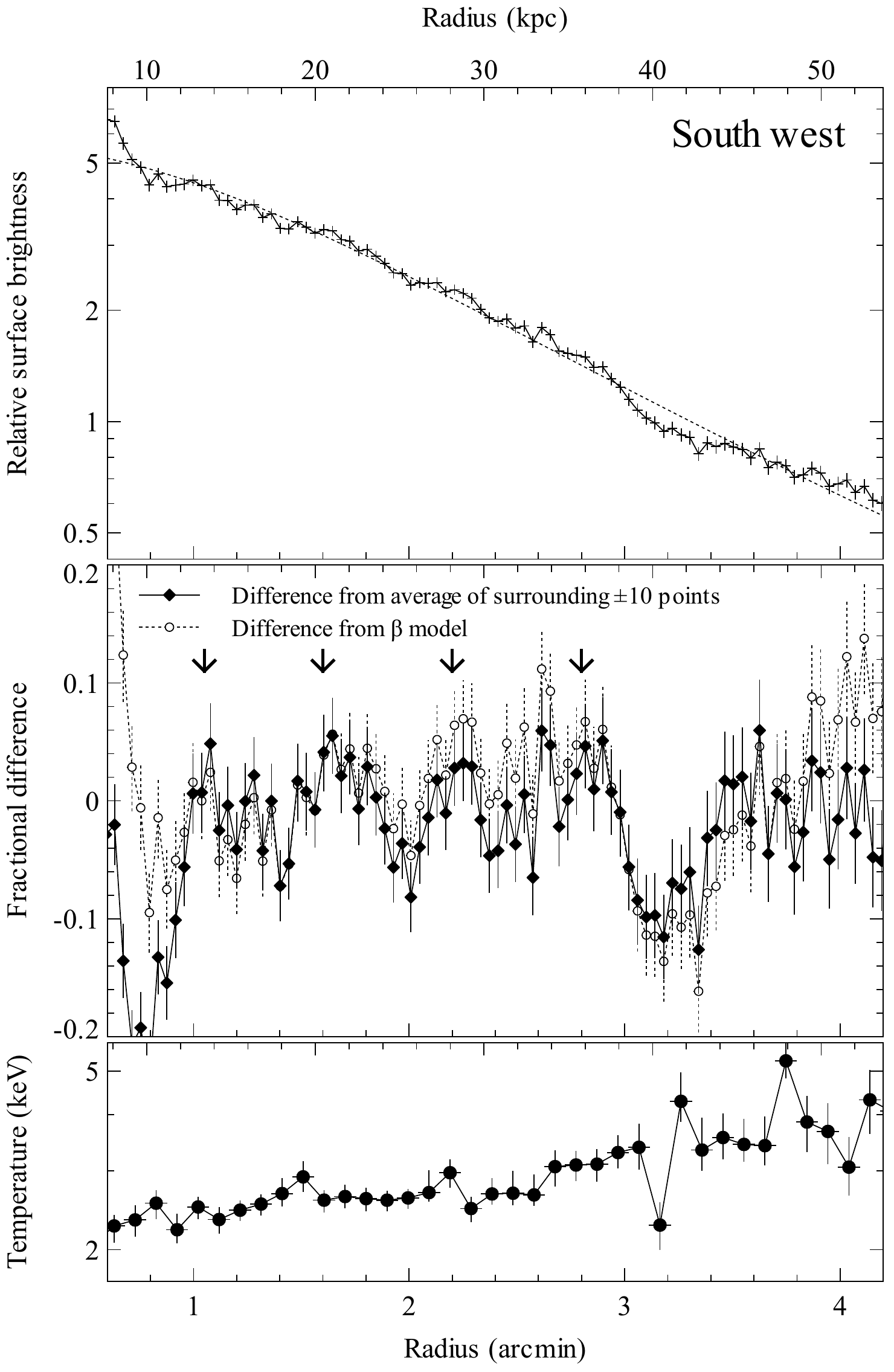}
  \caption{The top panel shows a surface brightness profile in a
    south-west sector with a $\beta$ model fit (dotted line). The
    centre panel shows the fractional residuals from this fit. It also
    shows the fractional residuals from the average of the 10 inner
    and outer radial bins for each bin. The arrows show the features
    mentioned in the text. The bottom panel shows projected
    temperatures determined from spectral fitting.}
  \label{fig:sector4}
\end{figure}

We plot the profile in the sector towards the south west in
Fig.~\ref{fig:sector4}. Shown in the top of the panel is a surface
brightness profile in 120 equal-width (0.04 arcmin) radial bins. We
also plot a $\beta$-model fit to the surface brightness profile in the
radial range shown. In the lower panel we show the fractional
residuals from the data to the model. In addition we show the
fractional residuals of the surface brightness of each bin to the
average of the 10 bins inside and outside of that bin (excluding bins
beyond the edge of the profile). The two different sets of residuals
are similar. We only show the residuals from the averaging technique
in later profiles here, as it is more difficult to fit a smooth
profile to these other sectors. The averaging method appears to be
robust. Note how the surface brightness significantly oscillates
around the zero value. There are strong positive features around 1.05,
1.6, 2.2, and 2.8~arcmin in radius (plotted as arrows). The feature in
the image at 3.6~arcmin is seen at low significance in this profile,
but only a small part of it is captured here. The large negative
feature at 3.2~arcmin is the sharp drop in surface brightness seen in
the right-hand panel in Fig.~\ref{fig:fftimage}. The amplitude of the
variation is typically of the order of 4 or 5 per~cent. An estimate
for the wavelength of the features is around 0.7~arcmin (9~kpc).

\begin{figure}
  \includegraphics[width=\columnwidth]{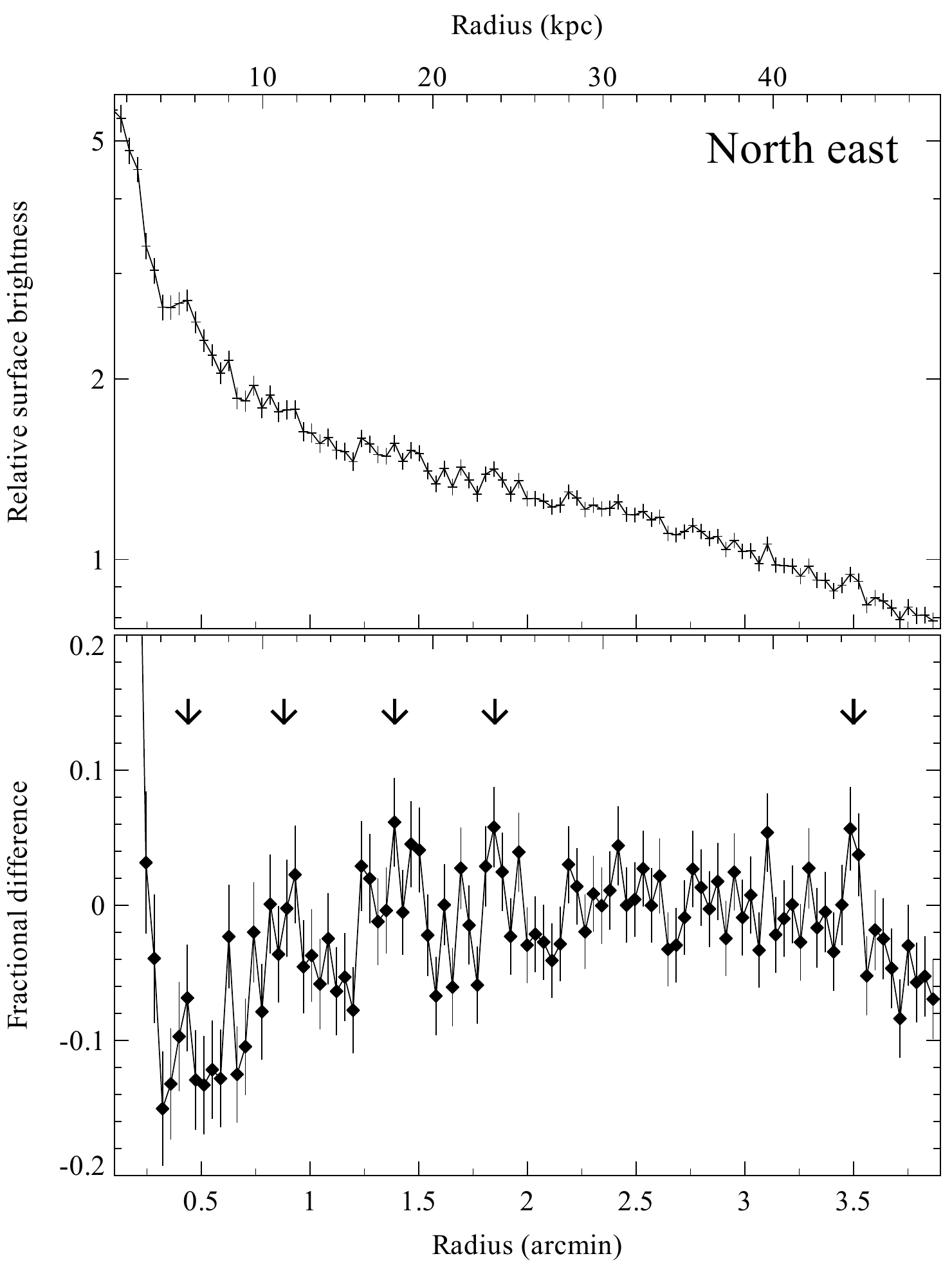}
  \caption{Surface brightness profile along the north east sector and
    fractional difference between each point and the average of the
    surrounding 10 inner and 10 outer radial bins. Features mentioned
    in the text are highlighted with arrows.}
  \label{fig:sector5}
\end{figure}

We show in Fig.~\ref{fig:sector5} the north east sector profile
generated in 100 equal-width (0.038 arcmin) bins and the fractional
residuals from the surrounding $\pm 10$ bins. In this image we see
positive features at 0.44, 0.88, 1.39, 1.85 and 3.5 arcmin. All of
these enhancements are seen in the FFT-filtered image. The inner peak
may be associated with the edge of the possible radio bubble seen as a
low pressure region by \cite{Crawford05}.

\begin{figure}
  \includegraphics[width=\columnwidth]{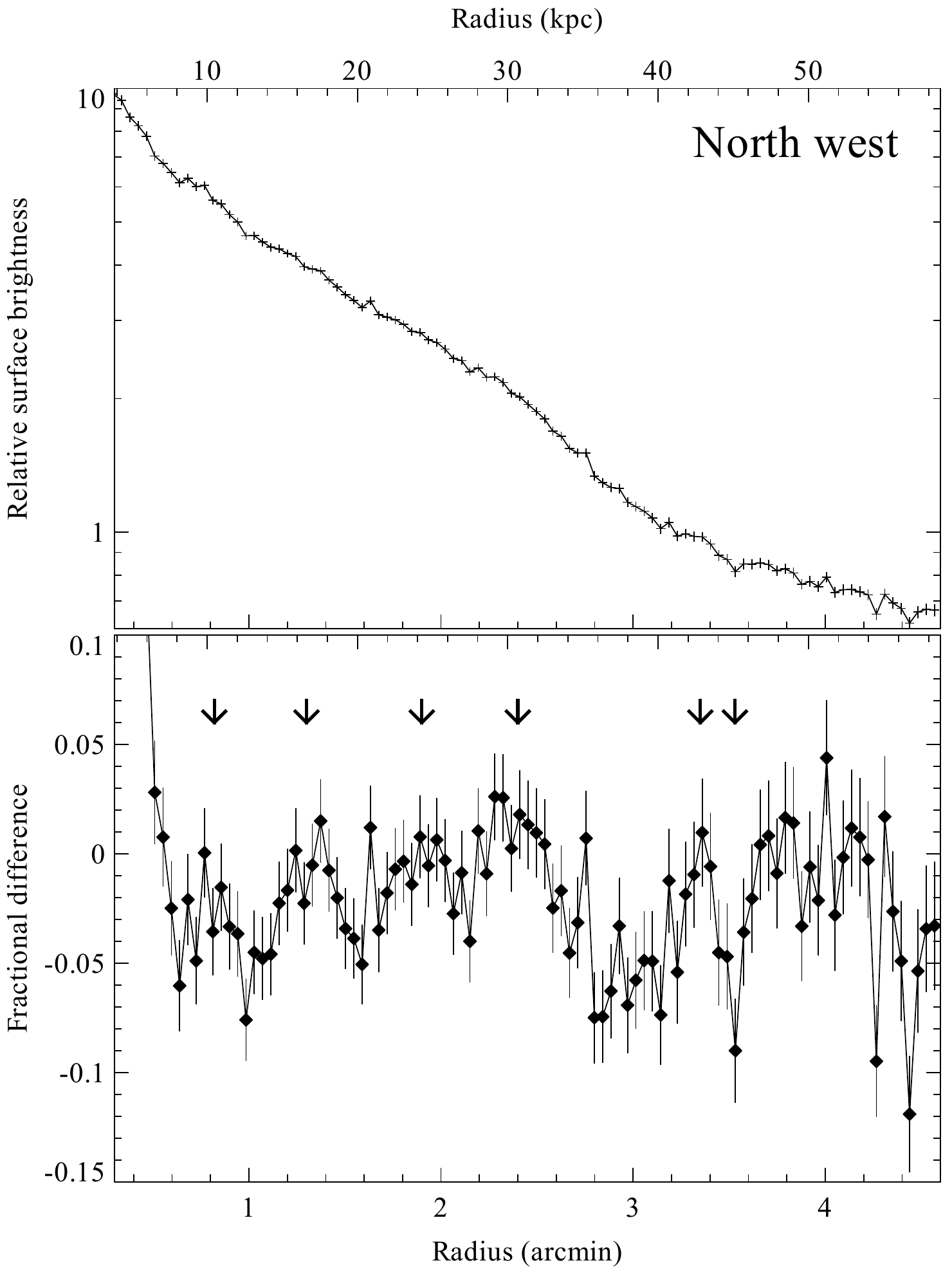}
  \caption{The surface brightness profile along a sector to the north
    west, plotted in the top panel. Again in the bottom panel we show
    the fractional difference between each point and the average of
    the surrounding 10 inner and outer radial bins.  The features
    mentioned in the text are highlighted with arrows.}
  \label{fig:sector7}
\end{figure}

Our final surface brightness profile is a sector towards the north
west of the cluster core (Fig.~\ref{fig:sector7}). There are around
four sets of obvious peaks in the inner part of the cluster, at radii
of 0.82, 1.3, 1.9 and 2.4 arcmin. There is also another peak at 3.35
arcmin, followed by a sharp drop (as seen in the FFT-filtered image)
at 3.53 arcmin, then another rise upwards. The size of the variations
in this sector appear smaller than the other two, with an amplitude of
around 3 per~cent.
  
\subsection{Temperature variations}
If the ripples are due to adiabatic sound waves, then temperature
fluctuations are expected at about one third of the relative amplitude
of the surface brightness variations. The amplitude of the temperature
fluctuations should therefore be 1--2 per cent in both the Perseus and
Centaurus clusters.  It is not clear whether this is consistent with
the range seen in the Perseus cluster
\citep{SandersPer07,Graham08Per}, although the multiphase nature of
the gas means that there are systematic uncertainties at the same
level which preclude any definitive measurement. For the Centaurus
cluster the uncertainties on the temperature are larger
(Fig.~\ref{fig:sector4}) and no detection is possible at this level
with current data.

\subsection{Projection effects}
To compute the real amplitude of the ripples in the absence of
projection effects in the Perseus cluster, we simulated projected
profiles of ripples with a known amplitude on an emissivity
model. Here we compare this method with another where we directly
deproject the surface brightness profile.

Starting from the outer radial bin in a sector, the surface brightness
is converted to an emissivity, assuming that all the emission comes
from a shell with those radii. The method examines successively
interior bins, subtracting off the contribution to the projected
emission from outer shells, converting the surface brightness to an
emissivity. To calculate the uncertainties on the emissivity profile a
Monte Carlo method is used. The input profiles are perturbed by
drawing from a Gaussian distribution using the best fitting values as
the central value, and uncertainties as widths. The deprojection is
repeated many times.  The emissivity from the median iteration and
84.2 and 15.8 percentiles are used to calculate the emissivity of a
bin and its uncertainty. The assumption of this method is that the
cluster is spherically symmetrical.

Using this deprojection technique, we obtain a ratio between the
projected surface brightness and the emissivity after deprojection of
around 4.  This is consistent with a factor of 5, computed using a
numerical model where we integrate a cluster emissivity model along
the line of sight (using a sinusoid with wavelength of 0.7~arcmin on a
power law emissivity profile with an index of -1.6). Therefore the
density variations are around 2.5 times the surface brightness
variations. 3 to 5 per cent surface brightness variations translate to
7.5 to 12.5 per cent density variations.

\section{Discussion}
Linear features in exposure-corrected images can be caused by
inaccurate exposure correction, for instance, bad pixels not properly
accounted for, or a variation of spectrum with source position. We
believe none of the features in the images presented are
exposure-correction artifacts, as the constituent exposure maps have
no features aligned in the same direction. Any artifacts in the images
should by aligned with the edge of the CCDs or the readout
direction. Most of the ripples also appear to have curvature, which
artifacts would not.

The Fourier filtering process can introduce spurious ripple-like
features. We have tested that the features are robust in the image
presented by trying different filtering levels. The surface brightness
profiles, which have no Fourier filtering applied, clearly demonstrate
that the ripples are real features.

Rather than pressure waves in the ICM, another mechanism which can
give surface brightness variations are fluctuations in the metallicity
of the ICM. The ICM can contain high metallicity blobs a few kpc in
size \citep{SandersPer07} and metallicity map of Centaurus is not
smooth \citep{Fabian05}. In order to check that metallicity is not the
cause of the observed ripples, we compared the spectra in the
south-western sector between radii of 1.53 and 1.86~arcmin, where
there is excess surface brightness, and 1.86 and 2.15~arcmin, where
there is a deficit. The spectra and the strength of the iron line were
very similar, with no significant differences in temperature or
metallicity measured by spectral fitting. In this case the region with
the brightness deficit was slightly higher in metallicity by $0.12 \pm
0.07 \Zsun$.  As a function of radius the brightness peaks correlate
with emission measure, not temperature or metallicity.  In neither the
Perseus nor Centaurus clusters do the observed metallicity variations
in a metallicity map correlate with the observed surface brightness
ripples.

To the west of the cluster at a radius of 30~kpc (2.3 arcmin), the
deprojected electron density is around $9\times 10^{-3} \pcmcu$ and
the temperature is 2.5~keV \citep{SandersReson06}. The mean radiative
cooling time, $t_\mathrm{cool}$, is 3.3~Gyr at this radius.  Using
equation (3) of \cite{SandersPer07}, the pressure ripples have a
height $\delta P \sim 7\times 10^{-3} \keV\pcmcu$ for 10 per cent
density ripples, assuming $\gamma=5/3$. 10 per cent density ripples
would correspond to a Mach number of 1.07.  The pressure amplitude
translates into a spherical sound wave power at this radius, $4 \pi
r^2 \delta P^2/(2 \rho c) = 5 \times 10^{42} \ergps$, were $r$ is the
radius, $\rho$ is the mass density and $c$ is the sound speed (the
factor of 2 converts from amplitudes to RMS energy). A 9~kpc
wavelength corresponds to a wave period of $10^7$~yr. This period is
close to that inferred for the ripples in the Perseus cluster
\citep{FabianPer03}.

If we choose a radius of 15~kpc instead (1.2 arcmin, where
$t_\mathrm{cool} = 1.4\Gyr$), the electron density is $1.5 \times
10^{-2} \pcmcu$ and the temperature is 2.0~keV. 10 per cent density
ripples translate into $1.5 \times 10^{42} \ergps$ of wave power.

If the inner features are 5 per cent surface brightness instead of 4,
the power is boosted by around 60 percent. The numbers are uncertain
by a factor of a few due to the assumptions of spherical symmetry and
the true projection factor. The ripples are not simple monochromatic
waves with a single wavelength.

The power in the sound waves can be compared to the bolometric
luminosity of the core of the cluster.  If we fit the \emph{Chandra}
spectrum from the inner 15 kpc of Centaurus with a model made up of
multiple temperature components of 0.5, 1, 2, 4 and 8~keV, we obtain a
bolometric X-ray luminosity of $8.6\times 10^{42} \ergps$, when
absorption is removed. From the inner 30 kpc, we obtain a luminosity
of $1.3 \times 10^{43} \ergps$.  Similar luminosities are obtained
from a grating spectrum of the core of the cluster
\citep{SandersRGS08}, although the geometry on the sky makes
comparisons difficult.  The luminosity of the core is just over twice
the wave power calculated above. Therefore the sound waves are close
to what is required to combat cooling in the core of the cluster.

If we assume that the two inner radio bubbles contain $4PV$ of energy
and that they deposit this energy over the period of the sound waves
($10^7$ yr), we estimate they have a total heating power of $\sim 2
\times 10^{43} \ergps$. This is four times the power we infer from the
ripples. This means that the conversion of jet power via bubbles into
sound waves is remarkably efficient ($\sim 0.25$). \cite{Graham08Per}
find that the energy in the high pressure region around the bubbles in
the Perseus cluster is about $3.5PV$. These observational results mean
that the energy content of the radio bubbles is close to the energy
injected by the nucleus, contrary to the interpretation by
\cite{Binney07} of the simulations of \cite{OmmaBinney04}, where only
about 10 per cent of the jet energy was found to go into the bubbles.

\section{Conclusions}
We detect surface brightness deviations in a \emph{Chandra} image of
the Centaurus cluster of galaxies. These features are between 3 and 5
per cent in size. Assuming that they are sound waves, the wave power
is approximately $5\times 10^{42}\ergps$ in the inner 30~kpc, with a
period of $10^7$ yr and a wavelength of 9~kpc. The power is close to
the X-ray luminosity of this central region. It is therefore
energetically possible that sound waves are the mechanism by which the
power of the central black hole is transmitted quasi-isotropically
through the cluster. Provided that the power in the waves is gradually
dissipated as heat, they provide the means by which radiative cooling
is balanced by heating in the inner cluster core.

\section*{Acknowledgements}
ACF acknowledges the Royal Society for support. 

\bibliographystyle{mnras}
\bibliography{refs}

\clearpage
\end{document}